\documentclass[aps,prd,twocolumn,superscriptaddress,nofootinbib,floatfix]{revtex4-2}
\usepackage{amsmath,amssymb,bm}
\usepackage{graphicx}
\usepackage{booktabs}
\usepackage{slashed}
\usepackage[colorlinks=true,linkcolor=blue,citecolor=blue,urlcolor=blue]{hyperref}

\newcommand{\Lbr}{\Lambda_{\rm br}}
\newcommand{\lh}{\hat\lambda}
\newcommand{\Neff}{N_{\rm eff}}
\newcommand{\dNeff}{\Delta N_{\rm eff}}
\newcommand{\Tdec}{T_{\rm dec}}
\newcommand{\GeV}{\,\mathrm{GeV}}
\newcommand{\TeV}{\,\mathrm{TeV}}
\newcommand{\MeV}{\,\mathrm{MeV}}

\begin{document}

\title{Majorana dark matter from a thick-to-thin brane junction:\\ Modal stability, real-mediator coannihilation, and a dark-radiation imprint}

\author{Francisco A. Brito}
\email{fabrito@df.ufcg.edu.br}
\affiliation{Departamento de F\'{\i}sica, Universidade Federal de Campina Grande, 58429-900 Campina Grande, Para\'{\i}ba, Brazil}

\date{\today}

\begin{abstract}
We show that the localized spectrum of a $D=7$ thick-to-thin brane junction can host a thermal Majorana dark-matter candidate. The anisotropy of the wall widths isolates a light junction fermion, which a four-dimensional Majorana mass splits into two nearby states. The factorized transverse profiles impose a modal selection rule on all cubic couplings, and together with the relaxation of the softly deformed background this makes the lighter state stable without an imposed dark parity. Because the localized mediator is a real scalar, identical Majorana particles annihilate in the $s$ wave, and the relic coupling becomes nearly independent of the Majorana mass. The junction sector is a Standard-Model singlet whose light chiral mode is a source of dark radiation. Current CMB data restrict its decoupling temperature and imply $0.05\lesssim\dNeff\lesssim0.076$ for dark-matter masses between $10\GeV$ and $1.5\TeV$, a range accessible to next-generation CMB surveys.
\end{abstract}

\maketitle

\section{Introduction}\label{sec:intro}

The localization of matter on defects is one of the oldest mechanisms by which a higher-dimensional theory can produce an effectively lower-dimensional spectrum. Fermionic zero modes bound to solitons already appear in the Jackiw--Rebbi mechanism~\cite{Jackiw:1975fn}. The domain-wall realization of Rubakov and Shaposhnikov provided the prototype for localizing four-dimensional matter in an extra dimension~\cite{Rubakov:1983bb}, and these ideas became part of the broader brane-world program~\cite{ArkaniHamed:1998rs,Antoniadis:1998ig,Randall:1999ee,Randall:1999vf}. For smooth (thick) defects the same logic extends beyond zero modes. The wall background defines a one-dimensional spectral problem whose bound states fix the localized masses, and whose overlap integrals fix the effective four-dimensional couplings~\cite{Bazeia:2008zx,Paulino:2022}.

Intersections of domain walls make this mechanism more effective because localization occurs independently along several transverse directions. In the thick-to-thin construction of Paulino and Brito~\cite{Paulino:2022}, an anisotropic hierarchy of wall widths produces a correspondingly hierarchical spectrum at a common $(3+1)$-dimensional junction. In the $D=7$, $N=3$ specialization used here, one thick wall and two thin walls isolate the first excitation along the thick direction, $F_{100}$, at a mass of order $k\Lbr$ with $k\ll1$. All other massive localized fermions lie near $\Lbr$. We ask whether this state can account for the dark matter.

Extra dimensions have long served as an organizing principle for dark sectors, from stable Kaluza--Klein states~\cite{Servant:2002aq} to warped constructions with fermionic dark matter~\cite{Vecchi:2013,Carmona:2020} and scenarios in which the size of a dark dimension sets the dark-sector scales~\cite{Anchordoqui:2023}. The candidate considered here is different. It is neither an independent bulk field with an imposed stabilizing symmetry nor a generic KK excitation. It descends from a particular localized junction mode, and the transverse wave functions that generate its mass hierarchy also restrict its cubic interactions.

To obtain a Majorana candidate we add a four-dimensional Majorana mass to the localized Dirac mode. This splits the Dirac level into two nearby Majorana eigenstates $X_\pm$, the pseudo-Dirac (inelastic) structure familiar from many dark-matter models~\cite{TuckerSmith:2001hy,DallaValleGarcia:2024}. The Majorana kernel is not derived from the wall geometry; it is an additional operator of the localized effective theory. The geometric content lies in the Dirac mass, the scalar and fermion wave functions, their overlap algebra, and the response of the localized levels to the soft deformation of the background.

Two features of the construction determine its thermal history.

\paragraph{(i) Reality of the localized mediator.} The scalar that couples $F_{100}$ to the light chiral mode $f\equiv F_{000}$ is the junction fluctuation $\sigma\equiv S_{100}$ of the real wall fields. For a Majorana fermion coupled to a light chiral fermion through a \emph{complex} (charged) $t$-channel scalar, annihilation into light fermions is helicity suppressed~\cite{Garny:2015wea,Ciafaloni:2011sa,Garny:2013ama,Arina:2020udz,Arina:2025whitepaper}. With a \emph{real} mediator, $f$-number is broken by the Majorana masses: $X_iX_j\to ff$ and $\bar f\bar f$ are allowed and have an unsuppressed $s$ wave, even for $i=j$. The mechanism is analogous to the lepton-number-violating annihilation $NN\to\nu\nu$ in the scotogenic model~\cite{Ma:2006km}, which arises from the mass splitting between the real and imaginary parts of the neutral inert scalar; here the mediator is real from the outset, and the effect is maximal. We compute these channels explicitly and show that they dominate the identical-particle rates (Sec.~\ref{sec:xsec}).

\paragraph{(ii) A self-contained thermal sector.} The three odd states $X_-$, $\sigma$, $X_+$ lie within a few GeV of each other. We include $\sigma$ as a coannihilating species and compute its annihilation exactly. It is $s$-wave but suppressed by a cancellation between the two Majorana exchanges. The entire junction sector $\{X_\pm,\sigma,f\}$ is a Standard-Model (SM) singlet, so no electroweak embedding of the localized vertex is required. Its contact with the visible plasma is described by a single decoupling temperature $\Tdec$, which determines both the relic coupling and the dark radiation carried by $f$. Recent CMB measurements of $\Neff$~\cite{Calabrese:2025ACT} restrict $\Tdec$ to a narrow range and lead to a definite prediction for $\dNeff$.

The coannihilation formalism~\cite{Griest:1990kh,Gondolo:1990dk,Edsjo:1997bg} assumes relative chemical equilibrium among the coannihilating states. When this fails, conversion-driven freeze-out or coscattering takes over~\cite{Garny:2017rxs,Garny:2018icg,Heisig:2024,Profumo:2025}. We verify that the conversions in our setup exceed the Hubble rate by about twelve orders of magnitude at freeze-out, so the single-equation treatment is justified. For freeze-out in a sector with its own temperature we follow Refs.~\cite{Feng:2008mu,Pospelov:2007mp}.

Our main quantitative results are as follows. For $k=1/40$, $\Lbr=1\TeV$ and $m_M=1\GeV$ the candidate has $M_-=42.3\GeV$, and $\Omega_Xh^2=0.120$ requires a localized coupling $g\simeq0.168$, i.e.\ $\lh\simeq4.5$. The Majorana mass must exceed $m_M^{\rm crit}\simeq0.66\GeV$ for stability, and above this value $\lh$ varies by less than $6\%$ for $m_M$ up to $10\GeV$. More generally, $g\simeq0.17\,(M_-/42\GeV)^{1/2}$ and $\lh\simeq4.5\,(\Lbr/{\rm TeV})^{1/2}$, independently of the thickness ratio, so that the coupling remains perturbative up to $M_-\simeq1.5\TeV$. Compatibility with the ACT DR6 bound on $\Neff$ requires the junction sector to share the temperature of the visible plasma down to $0.9\lesssim\Tdec\lesssim7\GeV$ at the benchmark, which gives $\dNeff\simeq0.064$--$0.076$, and disfavors masses below about $10\GeV$.

The paper is organized as follows. Section~\ref{sec:junction} reviews the junction spectrum and Sec.~\ref{sec:majorana} the Majorana deformation and localized couplings. Section~\ref{sec:stability} derives the soft level ordering and the modal stability criterion. Section~\ref{sec:xsec} gives the complete tree-level annihilation cross sections, and Sec.~\ref{sec:thermal} the thermal history. Sections~\ref{sec:relic} and~\ref{sec:neff} present the relic abundance and the dark-radiation prediction. Section~\ref{sec:pheno} discusses further phenomenology and the scope of the construction, and Sec.~\ref{sec:concl} concludes. The appendices contain the soft-splitting coefficient, the Majorana amplitudes, the modal-charge argument and the numerical procedure.

\section{The thick-to-thin junction and its localized spectrum}\label{sec:junction}

The construction is the $N=3$, $D=7$ specialization of the intersecting-wall mechanism of Ref.~\cite{Paulino:2022}. Three real scalar fields $\phi_i$ generate three codimension-one domain walls (5-branes), each depending predominantly on one transverse coordinate $x_i$. The walls intersect orthogonally on a $(3+1)$-dimensional junction. The background is generated by
\begin{equation}
W=\sum_{i=1}^3\lambda_i\left(\frac{\phi_i^3}{3}-a_i^2\phi_i\right),\qquad
V_\epsilon=\frac12\sum_i W_i^2+\frac{\epsilon}{2}\,F(\phi_i),
\label{eq:W}
\end{equation}
where $W_i=\partial W/\partial\phi_i$. The soft term proportional to $\epsilon$ couples the otherwise decoupled walls and controls the junction energetics, and the Yukawa operator $W_{\phi_i\phi_j}\bar\psi_i\psi_j$ localizes fermions. For small $\epsilon$ the zeroth-order kinks are $\phi_i^{(0)}=-a_i\tanh(\lambda_ia_ix_i)$. The thick-to-thin hierarchy is introduced through
\begin{equation}
a_1=k\,a,\qquad a_2=a_3=a,\qquad 0<k\ll1,
\label{eq:hier}
\end{equation}
so that the inverse widths are $\beta_1=k\Lbr$ and $\beta_{2,3}=\Lbr$, with $\Lbr=|\lambda|a$. In the separated-wall approximation the fermion wave functions factorize,
\begin{equation}
\Psi_{\bm n}\propto\chi^{(1)}_{n_1}(x_1)\,\chi^{(2)}_{n_2}(x_2)\,\chi^{(3)}_{n_3}(x_3),\quad \bm n=(n_1n_2n_3),
\end{equation}
and each one-dimensional problem is a P\"oschl--Teller well with two bound states,
\begin{equation}
f_0(u)=\frac{\sqrt3}{2}\,\mathrm{sech}^2u,\qquad f_1(u)=\sqrt{\frac32}\,\tanh u\;\mathrm{sech}\,u,
\label{eq:profiles}
\end{equation}
at $m^2=0$ and $m^2=3\beta^2$, followed by a continuum starting at $m^2=4\beta^2$. The localized $N=3$ spectrum is therefore
\begin{equation}
M^2_{n_1n_2n_3}=3\Lbr^2\left(k^2n_1+n_2+n_3\right),\qquad n_i\in\{0,1\}.
\label{eq:spectrum}
\end{equation}
For $k=1/40$ and $\Lbr=1\TeV$ the eight localized levels are listed in Table~\ref{tab:spectrum}. The first massive level, $F_{100}$, lies at $43.3\GeV$. The lowest continuum threshold, in the thick direction, is $M_{\rm cont}^{(1)}=2k\Lbr=50\GeV$, and all other massive localized states lie near $1.7$--$2.4\TeV$.

\begin{table}[t]
\centering
\begin{tabular}{cccc}
\toprule
mode & $M$ [GeV] & mode & $M$ [GeV]\\
\midrule
000 & 0 & 100 & 43.301\\
010 & 1732.051 & 001 & 1732.051\\
110 & 1732.592 & 101 & 1732.592\\
011 & 2449.490 & 111 & 2449.872\\
\bottomrule
\end{tabular}
\caption{Localized fermion levels of the $D=7$ junction for $k=1/40$, $\Lbr=1\TeV$, Eq.~\eqref{eq:spectrum}.}
\label{tab:spectrum}
\end{table}

\section{Majorana deformation and localized couplings}\label{sec:majorana}

\subsection{Pseudo-Dirac splitting}
Represent the massive mode $F_{100}$ by two left-handed Weyl fields $\xi$ and $\eta$ and add a four-dimensional Majorana mass,
\begin{equation}
\mathcal L_m=-m_D\,\xi\eta-\frac12\,m_M\left(\xi\xi+\eta\eta\right)+{\rm h.c.},
\label{eq:Lm}
\end{equation}
with $m_D=\sqrt3\,k\Lbr$.
The mass eigenstates are $\psi_+=(\xi+\eta)/\sqrt2$ and $\psi_-=i(\xi-\eta)/\sqrt2$, with positive masses
\begin{equation}
M_\pm=m_D\pm m_M .
\end{equation}
We denote by $X_\pm$ the corresponding four-component Majorana fields. The parameter $m_M$ is an effective-theory deformation; predicting it would require a specified Majorana kernel in the higher-dimensional theory, which we do not assume. We shall see that the relic coupling depends on $m_M$ only weakly (Sec.~\ref{sec:relic}).

\subsection{Overlap algebra}
The nonvanishing one-dimensional triple overlaps of the profiles~\eqref{eq:profiles} are
\begin{equation}
J_{000}=\!\int\!du\,f_0^3=\frac{2\sqrt3}{5},\quad J_{011}=\!\int\!du\,f_0f_1^2=\frac{\sqrt3}{5},
\label{eq:J}
\end{equation}
with $J_{101}=J_{110}=J_{011}$, while overlaps with an odd number of odd profiles vanish. For the product junction this becomes the modal selection rule
\begin{equation}
I_{\ell mn}\neq0\quad\Longleftrightarrow\quad \ell=m\oplus n ,
\label{eq:selection}
\end{equation}
where $\oplus$ is componentwise addition modulo two of the binary labels of the scalar mode $S_\ell$ and the fermion modes $F_m$, $F_n$. Restoring dimensions with $\beta_1=k\Lbr$ and $\beta_{2,3}=\Lbr$, each one-dimensional overlap scales as $\beta_i^{1/2}$. Defining the dimensionless higher-dimensional Yukawa normalization
\begin{equation}
\lh\equiv\lambda\,\Lbr^{3/2},\qquad [\lambda]=-3/2,
\end{equation}
the coupling of $S_{100}$ to $F_{100}$ and $F_{000}$ is
\begin{equation}
y_{100}=0.33255\,\sqrt k\,\lh .
\label{eq:y100}
\end{equation}

\subsection{Localized interaction in the mass basis}
The light mode $f\equiv F_{000}$ is a single chiral (Weyl) fermion, and $\sigma\equiv S_{100}$ is a real scalar. Since only one chirality of $f$ exists, the localized vertex is $y_{100}\,\sigma\,\bar f P_R F_{100}+{\rm h.c.}$ In the Majorana mass basis it reads
\begin{equation}
\mathcal L_{\rm int}=-\sigma\sum_{i=\pm}\bar f\left(g_iP_R+g_i^*P_L\right)X_i,
\label{eq:Lint}
\end{equation}
with $g_+=g$, $g_-=\pm ig$ and
\begin{equation}
g=\frac{y_{100}}{\sqrt2}=0.23515\,\sqrt k\,\lh\;\stackrel{k=1/40}{=}\;0.03718\,\lh .
\label{eq:g}
\end{equation}
In Eq.~\eqref{eq:Lint} $f$ is written as a massless Majorana spinor, equivalent to the Weyl field; the $P_R$ term creates $f$ and the $P_L$ term creates $\bar f$. The relative phase $i$ follows from the diagonalization of the pseudo-Dirac mass matrix; only $g_-^2=-g_+^2$ enters the observables below. The factor $\sqrt k$ is a geometrical suppression; the value of $\lh$ is not fixed by the geometry.

\section{Soft level ordering and modal stability}\label{sec:stability}

\subsection{Scalar--fermion splitting}
Write the relaxed background as $\phi_i=\phi_i^{(0)}+\epsilon\varphi_i+\mathcal O(\epsilon^2)$. The scalar fluctuation operator receives both the direct soft Hessian and the contribution induced by $\varphi_i$; the fermion operator receives only the latter. At $\epsilon=0$ the scalar and fermion operators of the 100 sector are isospectral, and at first order
\begin{equation}
\Delta M^2_{100}\equiv M^2_{S_{100}}-M^2_{F_{100}}=C(k)\,\epsilon\,\Lbr^2 ,
\label{eq:split}
\end{equation}
with $C(k)$ obtained by projecting the operator difference on the 100 profile (Appendix~\ref{app:Ck}). Representative values are given in Table~\ref{tab:Ck}.

\begin{table}[t]
\centering
\setlength{\tabcolsep}{4pt}
\begin{tabular}{cccccc}
\toprule
$k$ & 0.020 & 0.025 & 0.040 & 0.060 & 0.100\\
\midrule
$C(k)$ & 4.59357 & 4.54410 & 4.40062 & 4.22032 & 3.89461\\
\bottomrule
\end{tabular}
\caption{First-order soft-splitting coefficient of Eq.~\eqref{eq:split}, computed as described in Appendix~\ref{app:Ck} (numerical uncertainty below $10^{-5}$).}
\label{tab:Ck}
\end{table}

For $\epsilon<0$ the $S_{100}$ level lies below the unperturbed $F_{100}$ level. Denoting by $\delta m_F$ the common first-order shift of the fermionic level,
\begin{equation}
\begin{aligned}
M_{F_{100}}&=m_D+\delta m_F+\mathcal O(\epsilon^2),\\
M_\sigma&=m_D+\delta m_F+\frac{C(k)\,\epsilon\,\Lbr^2}{2m_D}+\mathcal O(\epsilon^2).
\end{aligned}
\end{equation}
The common shift cancels in every comparison below. We absorb it into $m_D$ and keep $M_\pm=m_D\pm m_M$.

\subsection{Localized decay channels}
\begin{figure}[t]
\centering
\includegraphics[width=\columnwidth]{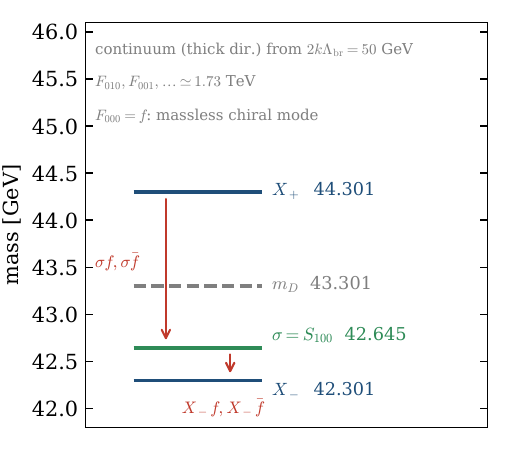}
\caption{Odd (modal charge $q_1=1$) sector of the junction at the benchmark~\eqref{eq:bench} with $m_M=1\GeV$. The heavier states decay sequentially, $X_+\to\sigma f,\sigma\bar f$ and $\sigma\to X_-f,X_-\bar f$, while $X_-\to\sigma f$ is closed because $m_M>m_M^{\rm crit}$. The light chiral mode $f$ and the TeV-scale junction modes are indicated schematically.}
\label{fig:spectrum}
\end{figure}

Starting from the modal label 100, a two-body decay into $F_{\bm n}$ requires the scalar $S_{100\oplus\bm n}$ by Eq.~\eqref{eq:selection}. The only localized fermion lighter than $F_{100}$ is $f=F_{000}$, so the only potentially open localized channel is $X_-\to f+\sigma$. The apparently more dangerous final state $F_{000}+S_{000}$ is forbidden by the selection rule. Channels involving $F_{010}$, $F_{001}$ or heavier modes are closed by their TeV-scale masses. The decay $X_-\to f\sigma$ is kinematically closed when $M_\sigma>M_-$, i.e. for
\begin{equation}
m_M>m_M^{\rm crit}=\frac{C(k)\,|\epsilon|\,\Lbr^2}{2m_D}=\frac{C(k)\,|\epsilon|}{2\sqrt3\,k}\,\Lbr,
\label{eq:mcrit}
\end{equation}
up to $\mathcal O(\epsilon^2)$ corrections. For the benchmark
\begin{equation}
k=\frac1{40},\qquad \Lbr=1\TeV,\qquad \epsilon=-0.02\,k^2,
\label{eq:bench}
\end{equation}
we obtain $m_M^{\rm crit}\simeq0.656\GeV$ (with $C(1/40)=4.5441$) at strict first order and $M_\sigma=42.645\GeV$. Using the unshifted fermion mass inside $\sqrt{m_D^2+C\epsilon\Lbr^2}$ would give $0.661\GeV$; this partially resums higher orders while omitting the common $\mathcal O(\epsilon)$ shift, so we do not use it. For $m_M>m_M^{\rm crit}$,
\begin{equation}
M_-<M_\sigma<M_+<M_{\rm cont}^{(1)}=50\GeV\ll M_{010,001},
\label{eq:ordering}
\end{equation}
as shown in Fig.~\ref{fig:spectrum}.

\subsection{Modal charge and the residual parity}
Assign to every localized mode the binary charge $q=\bm n\in(\mathbb Z_2)^3$. Equation~\eqref{eq:selection} implies $q_{\rm scalar}+q_{\rm fermion\,1}+q_{\rm fermion\,2}=0$ mod 2 at every retained cubic vertex. The Majorana mass~\eqref{eq:Lm} acts within the 100 sector and respects it. Summing the vertex constraints over a connected diagram cancels all internal charges pairwise, so the total external charge is conserved (Appendix~\ref{app:modal}). In particular, the first component $q_1$ acts as an exact parity of the retained localized theory under which $X_\pm$ and $\sigma$ are odd while $f$ is even. Every vertex of Eq.~\eqref{eq:Lint} contains exactly two odd fields. The lightest odd state, $X_-$, is therefore stable. The same parity forbids $X\sigma\to f\bar f$ at tree level and is the reason why only $XX$ and $\sigma\sigma$ pairs annihilate (Sec.~\ref{sec:xsec}).

The selection rule reflects the factorized structure of the background. The kink profiles are odd under $x_i\to-x_i$, and both $W_i^2$ and the soft term $F$ (Appendix~\ref{app:Ck}) are even in each $\phi_i$, so the combined reflection $x_i\to-x_i$, $\phi_i\to-\phi_i$ leaves the relaxed background invariant. The argument applies to the localized cubic interactions and is not an all-orders statement about an unspecified ultraviolet completion: operators that violate the factorization or this reflection symmetry would have to come from physics outside the junction construction.

\section{Complete tree-level annihilation}\label{sec:xsec}

\subsection{Role of the real mediator}
With a complex mediator carrying the charge of $f$, the interaction~\eqref{eq:Lint} would conserve $f$-number, and identical Majorana particles could only annihilate into $f\bar f$. In the limit $m_f\to0$ that channel has no $s$ wave~\cite{Garny:2015wea}. Here $\sigma$ is real, and the Majorana masses break $f$-number. The same vertex then produces $X_iX_j\to ff$ through two $P_R$ insertions and $X_iX_j\to\bar f\bar f$ through two $P_L$ insertions. A simple helicity argument shows that this channel is not suppressed. Two identical Majorana fermions at rest in an $s$ wave form the spin singlet $^1S_0$, with $J=0$. Two left-handed massless fermions emitted back to back have total angular-momentum projection $\lambda_1-\lambda_2=0$ along their axis, compatible with $J=0$. No chirality flip is needed, unlike for $f\bar f$, where $\lambda_1-\lambda_2=\pm1$. In the Dirac limit $m_M\to0$ these channels correspond to $DD\to ff$ and $\bar D\bar D\to\bar f\bar f$, allowed for a real mediator and absent for a complex one.

\subsection{Cross sections}
We computed all $2\to2$ amplitudes generated by Eq.~\eqref{eq:Lint} with the Feynman rules for Majorana fermions of Ref.~\cite{Denner:1992vza}, treating $f$ as a massless Majorana spinor so that $f$ and $\bar f$ are its two helicity states (Appendix~\ref{app:amp}). The spin sums were evaluated numerically with explicit spinors and integrated over angles. The thermal averages use the exact $\sigma(s)$. The leading terms of $\sigma v$, with $v$ the M{\o}ller velocity, are as follows.

\paragraph{$X_iX_j\to f\bar f$.}
For identical states the $s$ wave cancels and
\begin{equation}
\sigma v\,(X_iX_i\to f\bar f)=\frac{g^4M_i^2\,(M_i^4+M_\sigma^4)}{48\pi\,(M_i^2+M_\sigma^2)^4}\,v^2+\mathcal O(v^4),
\label{eq:pwave}
\end{equation}
the standard result for Majorana dark matter with a chiral $t$-channel scalar~\cite{Garny:2015wea}. For the mixed pair,
\begin{equation}
\sigma v\,(X_-X_+\to f\bar f)=\frac{g^4\,(M_-+M_+)^2}{32\pi\,(M_\sigma^2+M_-M_+)^2}+\mathcal O(v^2).
\label{eq:mixedffbar}
\end{equation}

\paragraph{$X_iX_j\to ff,\ \bar f\bar f$.} For all $i,j$, including $i=j$, the sum of the two channels is
\begin{equation}
\sigma v\,(X_iX_j\to ff,\bar f\bar f)=\frac{g^4\,(M_i+M_j)^2}{64\pi\,(M_\sigma^2+M_iM_j)^2}+\mathcal O(v^2),
\label{eq:ff}
\end{equation}
so that $\sigma v(X_iX_i\to ff+\bar f\bar f)=g^4M_i^2/[16\pi(M_i^2+M_\sigma^2)^2]$.

\paragraph{$\sigma\sigma\to ff,\ \bar f\bar f,\ f\bar f$.} Exchange of $X_\pm$ in the $t$ and $u$ channels gives
\begin{equation}
\begin{aligned}
&\sigma v\,(\sigma\sigma\to ff,\bar f\bar f)=\frac{g^4}{2\pi}\,\mathcal D^2+\mathcal O(v^2),\\
&\mathcal D=\frac{M_+}{M_\sigma^2+M_+^2}-\frac{M_-}{M_\sigma^2+M_-^2},
\end{aligned}
\label{eq:ss}
\end{equation}
where the relative sign is the one carried by $g_-^2=-g_+^2$. In the Dirac limit $\mathcal D\to0$, since a real scalar pair cannot produce two units of $f$-number when $f$-number is conserved. Moreover, $\partial_M[M/(M_\sigma^2+M^2)]$ vanishes at $M=M_\sigma$, so $\mathcal D$ is suppressed by the near-degeneracy of all three states. At the benchmark $\mathcal D=-8.1\times10^{-6}\GeV^{-1}$, and Eq.~\eqref{eq:ss} is about six orders of magnitude below Eq.~\eqref{eq:ff}. The $f\bar f$ channel of a real scalar pair is $d$-wave, $\sigma v\propto v^4$, as is known for real scalar dark matter with a fermionic $t$-channel mediator~\cite{Giacchino:2013bta}.

\paragraph{Absence of $X\sigma$ annihilation.} Every vertex contains two odd fields (Sec.~\ref{sec:stability}), so no tree-level diagram connects $X_i\sigma$ to a final state made only of $f$'s.

\paragraph{Checks.} The numerical amplitudes reproduce Eqs.~\eqref{eq:pwave} and~\eqref{eq:mixedffbar} to better than $10^{-6}$ when restricted to $f\bar f$. In the Dirac limit, $M_\pm\to M$ with $g=y/\sqrt2$, Eq.~\eqref{eq:mixedffbar} reduces to the standard chiral Dirac result $y^4M^2/[32\pi(M^2+M_\sigma^2)^2]$~\cite{Abercrombie:2015wmb}. The full effective rate at $n_+=n_-$ equals $\tfrac12(\sigma_{D\bar D}+\sigma_{DD})v$ with $\sigma_{DD}=\sigma_{D\bar D}$, as expected for a Dirac fermion coupled to a real scalar. The $\sigma\sigma\to ff$ amplitude vanishes to numerical precision for $M_+=M_-$.

\begin{table*}[t]
\centering
\begin{tabular}{lcc}
\toprule
initial state & dominant channel & $\sigma v$ at threshold [cm$^3$\,s$^{-1}$]\\
\midrule
$X_-X_-$ & $ff+\bar f\bar f$ ($s$) & $2.53\times10^{-26}$\\
$X_+X_+$ & $ff+\bar f\bar f$ ($s$) & $2.53\times10^{-26}$\\
$X_-X_+$ & $f\bar f$, $ff+\bar f\bar f$ ($s$) & $7.61\times10^{-26}$\\
$\sigma\sigma$ & $ff+\bar f\bar f$ ($s$, cancelling) & $9.7\times10^{-32}$\\
\bottomrule
\end{tabular}
\caption{Threshold annihilation rates at the benchmark point ($m_M=1\GeV$, $g=0.1679$). The $X_-X_+$ entry is the sum of Eqs.~\eqref{eq:mixedffbar} and~\eqref{eq:ff}; $f\bar f$ contributes two thirds of it.}
\label{tab:xsec}
\end{table*}

\section{Thermal history}\label{sec:thermal}

\subsection{Conversions and the single-equation reduction}
The ordering~\eqref{eq:ordering} allows the on-shell decays $X_+\to\sigma f,\sigma\bar f$ and $\sigma\to X_-f,X_-\bar f$. For massless $f$,
\begin{align}
\Gamma(\sigma\to X_-f)+\Gamma(\sigma\to X_-\bar f)&=\frac{g^2M_\sigma}{8\pi}\left(1-\frac{M_-^2}{M_\sigma^2}\right)^2,\\
\Gamma(X_+\to\sigma f)+\Gamma(X_+\to\sigma\bar f)&=\frac{g^2M_+}{16\pi}\left(1-\frac{M_\sigma^2}{M_+^2}\right)^2.
\end{align}
At the benchmark these are $1.2\times10^{-5}\GeV$ and $1.3\times10^{-4}\GeV$, while $H\simeq4\times10^{-18}\GeV$ at freeze-out. Since $f$ is relativistic and abundant, the inverse processes are equally fast. The three odd states therefore remain in relative chemical equilibrium throughout freeze-out, and after freeze-out every $\sigma$ and $X_+$ decays into $X_-$. Inverse decays $\sigma f\leftrightarrow X_\pm$ through both $f$ and $\bar f$ also relax any $f$-number chemical potential. Conversion-driven freeze-out~\cite{Garny:2017rxs} is not relevant here, and the total odd yield $Y=(n_-+n_++n_\sigma)/s$ obeys
\begin{equation}
\begin{aligned}
&\frac{dY}{dt}=-s\,\langle\sigma_{\rm eff}v\rangle\left(Y^2-Y_{\rm eq}^2\right),\\
&\langle\sigma_{\rm eff}v\rangle=\!\!\sum_{i,j\in\{-,+,\sigma\}}\!\! r_ir_j\,\langle\sigma_{ij}v\rangle,\quad r_i=\frac{n_i^{\rm eq}}{\sum_jn_j^{\rm eq}},
\end{aligned}
\label{eq:boltz}
\end{equation}
with $g_\pm=2$ and $g_\sigma=1$ internal degrees of freedom. The sum runs over ordered pairs, $\sigma_{\sigma X}=0$, and each $\langle\sigma_{ij}v\rangle$ is the exact relativistic thermal average~\cite{Gondolo:1990dk,Edsjo:1997bg}. Near freeze-out $(r_-,r_+,r_\sigma)\simeq(0.56,0.21,0.23)$ (Fig.~\ref{fig:channels}), so the scalar carries almost a quarter of the odd population. Since it annihilates only through the strongly suppressed rate~\eqref{eq:ss}, its presence dilutes $\langle\sigma_{\rm eff}v\rangle$.

\subsection{Junction-sector temperature and decoupling}
All fields of the junction sector, $\{X_\pm,\sigma,f\}$, are SM singlets. Their annihilation products are the light chiral modes $f$, which remain in the junction plasma. The only external input to the thermal history is how long this plasma shares the temperature of the visible sector. We parametrize it by a decoupling temperature $\Tdec$:
\begin{itemize}
\item for $T>\Tdec$ the two sectors share a common temperature, $T'=T$;
\item for $T<\Tdec$ the comoving entropies are separately conserved~\cite{Feng:2008mu},
\begin{equation}
\frac{s_{\rm D}(T')}{s_{\rm SM}(T)}=\frac{g_{*s}^{\rm D}(\Tdec)}{g_{*s}^{\rm SM}(\Tdec)}\equiv R ,
\label{eq:R}
\end{equation}
where $g_{*s}^{\rm D}$ counts $X_\pm$, $\sigma$ (with their full mass dependence) and $f$ ($7/8\times2$).
\end{itemize}
We solve Eq.~\eqref{eq:boltz} with $Y$ normalized to the total entropy, the equilibrium densities and thermal averages evaluated at $T'$, the Hubble rate including the junction-sector energy density, and a temperature-dependent SM $g_*(T)$ (Appendix~\ref{app:num}). The relic density is $\Omega_Xh^2=2.742\times10^8\,(M_-/\mathrm{GeV})\,Y_\infty\,(1+R)$.

The decoupling temperature parametrizes the contact between $f$ and the visible plasma. A natural source is a current--current operator $(\bar f\gamma^\mu P_Lf)(\bar\psi\gamma_\mu\psi)/\Lambda_p^2$, gauge invariant because $f$ is a singlet, generated by heavy junction or bulk states. By analogy with neutrino decoupling, which occurs near $1.5\MeV$ for $\Lambda\sim G_F^{-1/2}\simeq293\GeV$~\cite{Dolgov:2002wy}, such an operator decouples at $\Tdec\sim1.5\MeV\,(\Lambda_p/293\GeV)^{4/3}$. Decoupling at $\Tdec\sim1$--$7\GeV$ thus corresponds to $\Lambda_p\sim\mathcal O(10^2)\TeV$, well above $\Lbr$. A Higgs-portal coupling $|H|^2\sigma^2$, allowed by the residual parity, is an alternative. Our results depend on the portal only through $\Tdec$.

\section{Relic abundance}\label{sec:relic}

\begin{figure*}[t]
\centering
\includegraphics[width=0.48\textwidth]{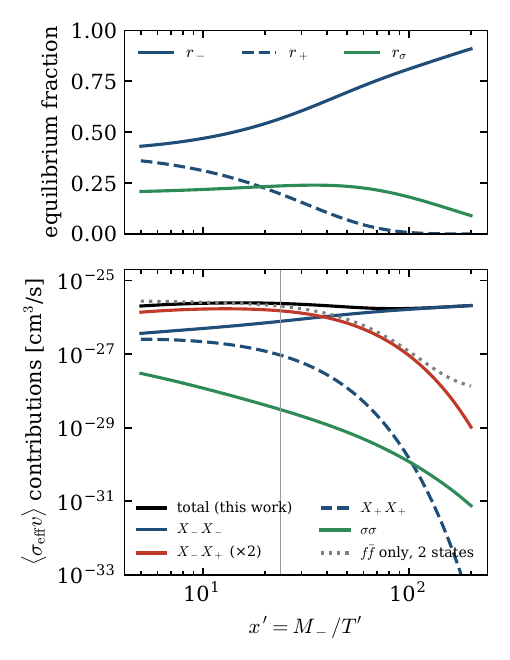}\hfill
\includegraphics[width=0.48\textwidth]{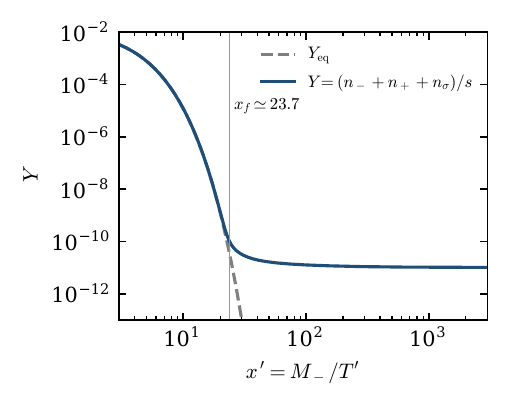}
\caption{Left: equilibrium fractions (top) and contributions $r_ir_j\langle\sigma_{ij}v\rangle$ to the effective rate (bottom) at the benchmark, $m_M=1\GeV$, $g=0.1679$, $\Tdec=1\GeV$. The dotted curve is the rate retained in a treatment with only the $f\bar f$ channels and only two coannihilating states, at the same coupling. The vertical line marks freeze-out. Right: evolution of the total odd yield.}
\label{fig:channels}
\end{figure*}

Figure~\ref{fig:channels} shows the individual contributions to the effective rate at the benchmark point. Before freeze-out the mixed $X_-X_+$ channel dominates. As $X_+$ and $\sigma$ become Boltzmann suppressed, the identical-particle channel $X_-X_-\to ff,\bar f\bar f$ takes over. Because that channel is $s$-wave, $\langle\sigma_{\rm eff}v\rangle$ stays nearly flat and annihilation continues efficiently after freeze-out. A treatment restricted to $f\bar f$ final states retains only the $p$-wave part, Eq.~\eqref{eq:pwave}, which falls by two orders of magnitude between $x'\sim20$ and $x'\sim200$.

\paragraph{Benchmark.} For the parameters~\eqref{eq:bench}, $m_M=1\GeV$ and $\Tdec=1\GeV$, the observed abundance $\Omega_ch^2\simeq0.120$~\cite{Aghanim:2018eyx} is obtained for
\begin{equation}
M_-=42.301\GeV,\qquad g=0.1679,\qquad \lh=4.51,
\label{eq:benchres}
\end{equation}
with freeze-out at $x_f'\simeq23.7$ ($T'_f\simeq1.8\GeV$).
The coupling is perturbative: $\lh>1$ only parametrizes the dimensionful higher-dimensional Yukawa, while the four-dimensional vertex is $g\simeq0.17$.

\begin{figure}[t]
\centering
\includegraphics[width=\columnwidth]{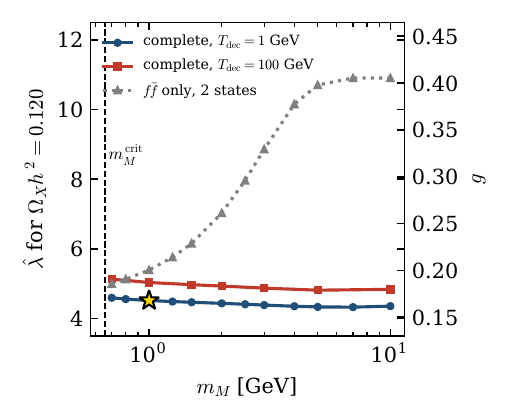}
\caption{Coupling required for $\Omega_Xh^2=0.120$ as a function of the Majorana mass, for the complete three-state calculation with $\Tdec=1\GeV$ (circles) and $\Tdec=100\GeV$ (squares). Triangles show a two-state calculation restricted to $f\bar f$ final states. The dashed line marks $m_M^{\rm crit}$ and the star the benchmark.}
\label{fig:relic}
\end{figure}

\paragraph{Dependence on $m_M$ and $\Tdec$.} Figure~\ref{fig:relic} and Table~\ref{tab:relic} show the coupling required for $\Omega_Xh^2=0.120$. With the complete set of channels the result depends only weakly on the Majorana mass: $\lh$ varies between $4.33$ and $4.59$ for $0.7\le m_M\le10\GeV$ when $\Tdec=1\GeV$. The $s$-wave identical-particle channel removes the strong dependence on the $X_+$ population that a $p$-wave-only treatment would imply. That treatment, shown dotted, would require $\lh$ to grow from $5.0$ to $10.9$ over the same range. This is relevant because $m_M$ is the only parameter of the dark sector that is not fixed by the geometry. If the junction sector decouples above the electroweak scale ($\Tdec=100\GeV$), it is hotter than the SM at freeze-out, and $\lh$ rises by about $7\%$.

The two new ingredients of the calculation act in opposite directions. Relative to a treatment restricted to $f\bar f$ final states, two states, a fixed $g_*=80$ and a common temperature, which gives $\lh=5.39$, adding the $ff,\bar f\bar f$ channels alone lowers the benchmark coupling to $\lh=3.97$. Including the $\sigma$ population, which dilutes the rate, and the temperature dependence of $g_*$ raises it to $4.51$.

\begin{table*}[t]
\centering
\begin{tabular}{ccccc}
\toprule
$m_M$ [GeV] & $M_-$ [GeV] & $\lh$ ($\Tdec=1$ GeV) & $g$ & $\lh$ ($\Tdec=100$ GeV)\\
\midrule
0.7 & 42.601 & 4.594 & 0.1708 & 5.129\\
1.0 & 42.301 & 4.515 & 0.1679 & 5.033\\
1.5 & 41.801 & 4.465 & 0.1660 & 4.970\\
2.0 & 41.301 & 4.434 & 0.1649 & 4.931\\
3.0 & 40.301 & 4.384 & 0.1630 & 4.871\\
5.0 & 38.301 & 4.331 & 0.1610 & 4.812\\
10.0 & 33.301 & 4.353 & 0.1619 & 4.835\\
\bottomrule
\end{tabular}
\caption{Relic-density solutions, $\Omega_Xh^2=0.120$, at $k=1/40$, $\Lbr=1\TeV$, $\epsilon=-0.02k^2$.}
\label{tab:relic}
\end{table*}

\begin{figure}[t]
\centering
\includegraphics[width=\columnwidth]{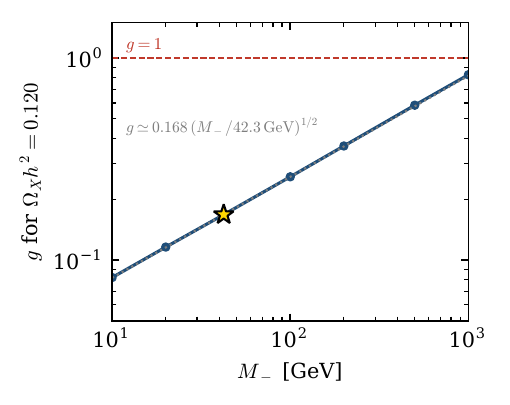}
\caption{Localized coupling required for $\Omega_Xh^2=0.120$ as a function of the dark-matter mass, at the mass ratios of the benchmark (points: full calculation; dotted: $M_-^{1/2}$ scaling). The star is the benchmark.}
\label{fig:mass}
\end{figure}

\paragraph{Other junction scales.} The benchmark uses $k=1/40$ and $\Lbr=1\TeV$. Since all masses of the odd sector scale with $m_D=\sqrt3\,k\Lbr$ at fixed $m_M/m_D$ and $\epsilon/k^2$, the cross sections scale as $g^4/M_-^2$, and the relic condition becomes a relation between $g$ and $M_-$. Repeating the full calculation for $10\GeV\le M_-\le1\TeV$, with $\Tdec=\max(1\GeV,M_-/42)$, gives (Fig.~\ref{fig:mass})
\begin{equation}
g\simeq0.168\left(\frac{M_-}{42.3\GeV}\right)^{1/2},\qquad
\lh\simeq4.5\left(\frac{\Lbr}{1\TeV}\right)^{1/2},
\label{eq:scaling}
\end{equation}
to within $2\%$ over the whole range. The second relation follows from Eq.~\eqref{eq:g}, $g\propto\sqrt k\,\lh$, together with $M_-\propto k\Lbr$: the required higher-dimensional coupling is independent of the thickness ratio $k$ and is fixed by the brane scale alone. Perturbativity of the four-dimensional vertex, $g\lesssim1$, limits the candidate to $M_-\lesssim1.5\TeV$.

\paragraph{Uncertainties.} The leading theoretical uncertainty is the SM equation of state near $T\sim2\GeV$. We use an ideal-gas $g_*(T)$ with physical masses; perturbative and lattice corrections modify WIMP abundances at the $\lesssim10\%$ level in this mass range~\cite{Saikawa:2020swg}. Since $\Omega\propto g^{-4}$ at fixed thermal history, this translates into an uncertainty of about $2$--$3\%$ in $\lh$. Higher-order corrections in $\epsilon$ shift $m_M^{\rm crit}$ but not the relic coupling at the conservative benchmark, which lies $0.34\GeV$ above the threshold. Radiative processes, such as internal bremsstrahlung, are subleading once the two-body rate is unsuppressed.

\section{Dark radiation}\label{sec:neff}

\begin{figure}[t]
\centering
\includegraphics[width=\columnwidth]{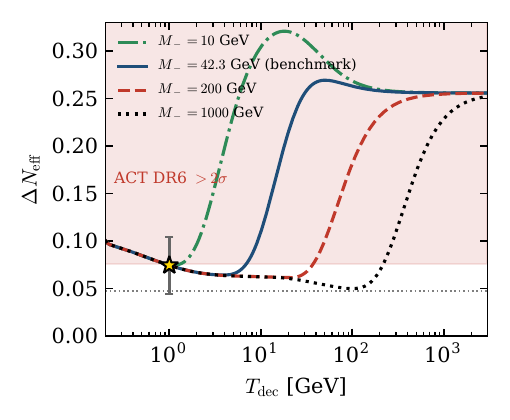}
\caption{Dark radiation carried by the chiral junction mode $f$ as a function of the decoupling temperature. The shaded region exceeds the ACT DR6 (P-ACT-LB) $2\sigma$ upper bound~\cite{Calabrese:2025ACT}. Curves correspond to different dark-matter masses, at the mass ratios of the benchmark. The star is the benchmark, with an error bar of $\sigma(\Neff)=0.03$ indicating the reach of next-generation surveys. The dotted line is the minimal contribution, $0.047$, of a Weyl fermion that decouples above all SM thresholds.}
\label{fig:neff}
\end{figure}

After the odd states have annihilated, the entire entropy of the junction sector resides in the massless chiral mode $f$. It constitutes a dark-radiation component whose temperature follows from Eq.~\eqref{eq:R}. Before $e^\pm$ annihilation $g_{*s}^{\rm SM}=10.75$ and $T_\nu=T$, so
\begin{equation}
\dNeff=\left(\frac{T_f}{T_\nu}\right)^4=\left[\frac{10.75}{7/4}\;\frac{g_{*s}^{\rm D}(\Tdec)}{g_{*s}^{\rm SM}(\Tdec)}\right]^{4/3}.
\label{eq:dneff}
\end{equation}
Two regimes follow (Fig.~\ref{fig:neff}).
\begin{itemize}
\item If the sector decouples while $X_\pm$ and $\sigma$ are relativistic, $\Tdec\gtrsim M_-$, their entropy is later transferred to $f$ alone. With $g_{*s}^{\rm D}=6.25$ this gives $\dNeff\simeq0.26$. This is about $5.5$ times the minimal contribution, $0.047$, of a Weyl fermion that decouples above all SM thresholds~\cite{Trendafilova:2026}.
\item If the sector shares the SM temperature until the odd states are Boltzmann suppressed, $\Tdec\ll M_-$, then $g_{*s}^{\rm D}(\Tdec)=7/4$ and $\dNeff=[10.75/g_{*s}^{\rm SM}(\Tdec)]^{4/3}$. This lies between $0.064$ and $0.08$ for $0.7\lesssim\Tdec\lesssim7\GeV$ and grows rapidly below the QCD crossover.
\end{itemize}
At the benchmark, $\Tdec=1\GeV$ and $\dNeff=0.074$.

\paragraph{Comparison with data.} The ACT DR6 analysis combined with Planck, CMB lensing and DESI BAO gives $\Neff=2.86\pm0.13$, or $2.89\pm0.11$ with primordial abundances~\cite{Calabrese:2025ACT}. With the Standard-Model value $3.044$~\cite{Bennett:2020zkv}, the early-decoupling branch, $\Neff\simeq3.30$, lies about $3.4\sigma$ above the measured central value and is disfavored. The window that survives at the $2\sigma$ level, $\dNeff<0.076$ with respect to the Gaussian P-ACT-LB result, is
\begin{equation}
0.9\lesssim\frac{\Tdec}{\rm GeV}\lesssim7\ \Longrightarrow\ 0.064\lesssim\dNeff\lesssim0.076 .
\label{eq:window}
\end{equation}
The measured central value lies below the SM prediction itself, which is $1.4\sigma$ high. The benchmark, $\Neff=3.118$, adds only $0.6\sigma$ to this offset. Within the window $\lh$ changes by less than $0.3\%$ (Table~\ref{tab:relic} and Sec.~\ref{sec:relic}), so the relic prediction and the $\Neff$ prediction are essentially independent. The upper edge of the window is set by the residual entropy of $X_\pm$ and $\sigma$ at decoupling, and the lower edge by the QCD crossover.

For other masses the window changes as shown in Fig.~\ref{fig:neff}. Its lower edge, set by the QCD crossover, is universal, while its upper edge grows with $M_-$ because the odd states become nonrelativistic at higher temperatures. For $M_-=10\GeV$ the window shrinks to $1.0\lesssim\Tdec\lesssim1.3\GeV$, so lighter candidates are disfavored. For $M_-=200\GeV$ and $1\TeV$ it extends to $\Tdec\simeq36\GeV$ and $210\GeV$, and the smallest attainable values are $\dNeff\simeq0.061$ and $0.050$. Across the whole perturbative mass range the model therefore predicts $0.05\lesssim\dNeff\lesssim0.076$.

The prediction~\eqref{eq:window} can be tested directly. Next-generation CMB surveys designed for $\sigma(\Neff)\simeq0.03$ would detect $\dNeff\simeq0.07$ at the $2\sigma$ level~\cite{Abazajian:2019eic,Trendafilova:2026}. Should a future measurement exclude $\dNeff\gtrsim0.06$, the minimal junction sector would be ruled out unless $f$ acquires a mass, or a further light state alters the entropy budget. An additional massless real scalar in the junction sector would raise the prediction by a factor $11/7$ for $\Tdec\ll M_-$ (Sec.~\ref{sec:pheno}).

\section{Further phenomenology and scope}\label{sec:pheno}

\paragraph{Direct and indirect detection.} Since $X_-$ couples only to junction-sector fields, there is no tree-level coupling to nucleons. Scattering arises only through the portal that fixes $\Tdec$, at one loop and suppressed by $\Lambda_p^{-2}\sim(10^2\TeV)^{-2}$, far below current sensitivity. Present-day annihilation proceeds into $ff$ and $\bar f\bar f$, which are invisible, so there are no gamma-ray or cosmic-ray signals at tree level. The model is a secluded thermal relic in the sense of Refs.~\cite{Pospelov:2007mp,Feng:2008mu}, with the junction geometry supplying both the spectrum and the stabilizing selection rule.

\paragraph{Kinetic decoupling.} $X_-$ stays in kinetic equilibrium with $f$ through $X_-f\to X_-f$ scattering mediated by the nearby $\sigma$, with $M_\sigma-M_-=0.34\GeV$. For $E_f\ll M_\sigma-M_-$ the momentum-transfer rate scales as $g^4T'^6/[M_-(M_\sigma-M_-)^2]$ (up to numerical factors), which gives an estimated kinetic-decoupling temperature of $\mathcal O(1$--$10)\MeV$. The corresponding cutoff in the matter power spectrum lies at subsolar halo masses, far below observable scales.

\paragraph{Colliders.} The junction-sector states are SM singlets with $\mathcal O(10^2)\TeV$ portals, and the other junction modes sit at $\sim1.7\TeV$. Collider signals of the dark sector are therefore model dependent and not required for consistency; they would arise only if the heavy junction states that generate the portal are accessible.

\paragraph{Scalar modes of the junction.} We have followed Ref.~\cite{Paulino:2022} in labeling the localized scalar fluctuations $S_\ell$ by the same factorized profiles as the fermions, and in projecting the soft splitting on $f_1(kx_1)f_0(x_2)f_0(x_3)$. In the strictly separated-wall limit the fluctuation of $\phi_1$ is governed along $x_1$ by the same P\"oschl--Teller operator as the fermions, while its confinement along the thin directions relies on the coupling of the three walls at the junction through the soft term. A complete solution of the coupled scalar fluctuation problem at finite $\epsilon$ would modify the profiles beyond the factorized approximation. It would not alter the selection rule, which depends only on the parity of the profiles under $x_i\to-x_i$, a symmetry the coupled problem shares. It could change $C(k)$, but a $3\%$ change of $C$ shifts the relic coupling by only $0.2\%$. A second issue is the all-zero scalar mode $S_{000}$. If a massless scalar remains localized on the junction and shares the temperature of $f$, it adds to the dark radiation: for $\Tdec\ll M_-$ the prediction becomes $\dNeff\simeq0.10$--$0.12$, which ACT DR6 disfavors at more than $2\sigma$. Current CMB data therefore also constrain the scalar spectrum of the junction, and favor configurations in which the soft deformation lifts any localized scalar zero mode above the MeV scale.

\paragraph{Geometric and non-geometric inputs.} The Dirac mass, the profiles, the overlap algebra, the selection rule~\eqref{eq:selection}, the coefficient $C(k)$ and the reality of the mediator follow from the junction. The Majorana mass $m_M$ and the normalization $\lh$ do not. The relic coupling depends weakly on $m_M$ above threshold and fixes $\lh\simeq4.3$--$4.6$. The Majorana mass could arise from a dark Higgs mechanism, although a real scalar suffices to generate a pseudo-Dirac splitting~\cite{DallaValleGarcia:2024}; we do not pursue a gauged extension here, since it would add free parameters and light states that modify both the relic density and $\dNeff$. A higher-dimensional origin of the Majorana kernel, and a microscopic derivation of the portal and of the masses of any additional light junction scalars, are left for future work. None of them affects the stability mechanism or the annihilation channels computed here.

\section{Conclusions}\label{sec:concl}

We have shown that the thick-to-thin junction can host a complete thermal Majorana dark sector without an independent flavor ansatz or an imposed dark parity. The anisotropic $D=7$ geometry isolates the mode $F_{100}$. A four-dimensional Majorana mass produces the nearby states $X_\pm$. The factorized wave functions impose the modal selection rule $\ell=m\oplus n$, and the relaxed soft background orders the scalar $S_{100}$ above $X_-$ once $m_M>m_M^{\rm crit}$. Within the localized theory, $X_-$ is the lightest state carrying an exactly conserved modal parity.

The reality of the localized mediator changes the thermal phenomenology qualitatively. Fermion number of the light chiral mode is broken, identical Majorana particles annihilate in the $s$ wave into $ff$ and $\bar f\bar f$, and the relic coupling becomes almost independent of the Majorana splitting. The three odd states remain in chemical equilibrium through freeze-out, and the scalar annihilation, which we computed exactly, is suppressed by a pseudo-Dirac cancellation. No ingredient of the thermal history is left unspecified apart from $\Tdec$. For $k=1/40$ and $\Lbr=1\TeV$ the observed abundance requires $M_-\simeq42\GeV$ and $g\simeq0.16$--$0.17$, i.e. $\lh\simeq4.3$--$4.6$; more generally $\lh\simeq4.5\,(\Lbr/{\rm TeV})^{1/2}$, independently of the thickness ratio.

Because the junction sector is an SM singlet, its main cosmological signature is the dark radiation carried by $f$. Current CMB data require the sector to share the SM temperature until a few GeV, disfavor candidates lighter than about $10\GeV$, and leave $0.05\lesssim\dNeff\lesssim0.076$ over the perturbative mass range ($0.064$--$0.076$ at the benchmark), within reach of next-generation CMB experiments. The scenario can therefore be tested by cosmological observations.

\begin{acknowledgments}
This program was initiated in Ref.~\cite{Paulino:2022}. Part of the present investigation was presented at the Encontro de Primavera da Sociedade Brasileira de F\'{\i}sica, held in Vit\'oria, ES, Brazil, in September 2026. I am happy to thank all those who discussed specific aspects of this work with me, especially Prof.\ Nelson R. F. Braga (UFRJ) and Prof.\ Clarissa Siqueira (ON), as well as the organizers for a very fruitful meeting. The author acknowledges CAPES and CNPq for partial financial support, in particular CNPq Grant No.~309092/2022-1, and the Universal-FAPESQ-PB program.

\emph{Use of AI tools.} AI language models (ChatGPT, Grok and Claude) were used to assist in this work: in drafting and editing the text, in checking the literature, and in writing and cross-checking the numerical codes for the scattering amplitudes, thermal averages, Boltzmann evolution and soft-splitting coefficient. The physical setup, the choice of methods and the interpretation of the results are the author's. All analytical results and numerical values reported here were verified by the author, who takes full responsibility for the content of the article.
\end{acknowledgments}

\section*{Data availability}
The numerical codes used for the amplitudes, the thermal averages, the Boltzmann evolution and the soft-splitting coefficient, together with the data underlying the figures, are available from the author upon reasonable request.

\appendix

\section{First-order soft splitting and \texorpdfstring{$C(k)$}{C(k)}}\label{app:Ck}
The coefficient $C(k)$ of Eq.~\eqref{eq:split} follows from first-order perturbation theory around the relaxed background
\begin{equation}
\phi_i=\phi_i^{(0)}+\epsilon\,\varphi_i+\mathcal O(\epsilon^2),\qquad \phi_i^{(0)}=-a_i\tanh(\lambda_ia_ix_i).
\end{equation}
Linearizing the background equations gives
\begin{equation}
H_{B,i}^{(0)}\varphi_i=-\frac12\,F_i[\Phi^{(0)}],
\label{eq:lin}
\end{equation}
with $H^{(0)}_{B,i}$ the undeformed scalar fluctuation operator. For
\begin{equation}
F_{ij}=\frac12\left(\phi_i^4+\phi_j^4\right)-3\phi_i^2\phi_j^2+\frac92,
\end{equation}
the derivatives entering the fluctuation problem are
\begin{equation}
\begin{aligned}
F_i&=4\phi_i^3-6\phi_i\left(\phi_j^2+\phi_k^2\right),\\
F_{ii}&=6\sum_{j\ne i}\left(\phi_i^2-\phi_j^2\right),\quad F_{ij}=-12\phi_i\phi_j\ (i\ne j).
\end{aligned}
\end{equation}
The fermionic chirality whose undeformed operator is isospectral to the scalar one is $H_F=-\partial_i^2+G_i^2+G_i'$, with $G_i=W_{\phi_i\phi_i}=2\lambda_i\phi_i$. The relaxed background shifts it by
\begin{equation}
\delta H_F=\epsilon\left[8\lambda_i^2\phi_i^{(0)}\varphi_i+2\lambda_i\partial_i\varphi_i\right].
\end{equation}
Subtracting the two operators, rather than assigning the scalar correction to an absolute mass, gives the splitting,
\begin{equation}
\delta(H_B-H_F)_{ii}=\epsilon\left[4\lambda_i^2\phi_i^{(0)}\varphi_i-2\lambda_i\partial_i\varphi_i\right]+\frac{\epsilon}{2}F_{ii}.
\end{equation}
We work in units $\Lbr=\lambda=a=1$, so that $\phi_1^{(0)}=-k\tanh(kx_1)$ and $\phi_{2,3}^{(0)}=-\tanh x_{2,3}$, and we take $F=\sum_{i<j}F_{ij}$. The splitting is the expectation value of $\delta(H_B-H_F)_{11}$ in the normalized $100$ profile $\Psi_{100}=\sqrt k\,f_1(kx_1)f_0(x_2)f_0(x_3)$:
\begin{equation}
\begin{aligned}
C(k)&=-2R_{100}(k)+\frac65\left(3k^2-1\right),\\
R_{100}&=\left\langle\partial_1\varphi_1-2\phi_1^{(0)}\varphi_1\right\rangle_{100}.
\end{aligned}
\end{equation}
The second term is the direct soft Hessian: using $\langle\tanh^2\rangle_{f_1}=3/5$ and $\langle\tanh^2\rangle_{f_0}=1/5$ one finds $\tfrac12\langle F_{11}\rangle_{100}=\tfrac65(3k^2-1)$ exactly.

\paragraph{Decomposition of the source.} Writing $\phi_{2,3}^2=1-\mathrm{sech}^2x_{2,3}$, the source of Eq.~\eqref{eq:lin} for $i=1$ splits into an asymptotic part, which depends on $x_1$ only, and two identical pieces localized at the thin walls:
\begin{equation}
-\frac12F_1=\left[-2\phi_1^3+6\phi_1\right]-3\phi_1\left[\mathrm{sech}^2x_2+\mathrm{sech}^2x_3\right].
\end{equation}
By linearity, $\varphi_1=\varphi_\infty(x_1)+\psi(x_1,x_2)+\psi(x_1,x_3)$, and correspondingly
\begin{equation}
R_{100}=R^{(\infty)}_{100}+2R_q ,
\label{eq:Rsplit}
\end{equation}
where $R_q$ is the contribution of a \emph{single} thin direction. Both sources are odd in $x_1$, so the solutions lie in the odd sector of the operator $-\partial_1^2+k^2(6\tanh^2kx_1-2)$, which is gapped by $3k^2$; the even translational zero mode $\mathrm{sech}^2kx_1$ decouples. Only this odd sector contributes to the projection on the $100$ profile.

\paragraph{Asymptotic part.} The ordinary differential equation for $\varphi_\infty$ gives $R^{(\infty)}_{100}=-3+\tfrac45k^2$, reproduced numerically to $10^{-5}$.

\paragraph{Localized part.} Fourier transforming in $x_2$, with $\int dx\,\mathrm{sech}^2x\,e^{-iqx}=\pi q/\sinh(\pi q/2)$, reduces the two-dimensional problem to a family of ordinary equations,
\begin{equation}
\begin{aligned}
&\left[-\partial_1^2+k^2\left(6\tanh^2kx_1-2\right)+q^2\right]\tilde\psi_q(x_1)\\
&\qquad=-3\phi_1^{(0)}(x_1)\,\frac{\pi q}{\sinh(\pi q/2)},
\end{aligned}
\end{equation}
which we solve by finite differences with $\tilde\psi_q$ odd in $x_1$. The projection on $f_0(x_2)^2=\tfrac34\mathrm{sech}^4x_2$ is performed in Fourier space, using $\int dx\,\mathrm{sech}^4x\,e^{-iqx}=\pi q(q^2+4)/[6\sinh(\pi q/2)]$, and the $q$ integral by Gauss--Legendre quadrature.

\paragraph{Results.} Table~\ref{tab:Ckapp} lists the three contributions. At $k=0.025$ one has $R_q=0.064287$ per thin direction, $R_{100}=-2.99950+2\times0.064287=-2.87093$ and $C=4.54410$. The decomposition~\eqref{eq:Rsplit} makes explicit the factor of two associated with the two equivalent thin walls. The values are stable to below $10^{-5}$ under refinement of the grid ($\Delta u=5\times10^{-3}\to2.5\times10^{-3}$), the box size ($40\to60$ widths) and the number of Fourier nodes ($160\to240$). The calculation is first order in $\epsilon$; for $|\epsilon|=0.02k^2$, higher-order corrections lie outside the accuracy claimed.

\begin{table*}[t]
\centering
\begin{tabular}{cccccc}
\toprule
$k$ & $R^{(\infty)}_{100}$ & $R_q$ (per thin wall) & $R_{100}$ & $\tfrac65(3k^2-1)$ & $C(k)$\\
\midrule
0.020 & $-2.99968$ & 0.051808 & $-2.89607$ & $-1.19856$ & 4.59357\\
0.025 & $-2.99950$ & 0.064287 & $-2.87093$ & $-1.19775$ & 4.54410\\
0.040 & $-2.99872$ & 0.100645 & $-2.79743$ & $-1.19424$ & 4.40062\\
0.060 & $-2.99712$ & 0.146722 & $-2.70368$ & $-1.18704$ & 4.22032\\
0.100 & $-2.99200$ & 0.231349 & $-2.52930$ & $-1.16400$ & 3.89461\\
\bottomrule
\end{tabular}
\caption{Contributions to the soft-splitting coefficient, Eqs.~\eqref{eq:Rsplit} and following.}
\label{tab:Ckapp}
\end{table*}

\section{Majorana amplitudes}\label{app:amp}
We write $f$ as a massless Majorana spinor $\mathcal F$ and use $\mathcal L_{\rm int}=-\sigma\sum_i\bar{\mathcal F}\Gamma_iX_i$ with $\Gamma_i=g_iP_R+g_i^*P_L$. Since $C\Gamma_i^TC^{-1}=\Gamma_i$, the vertex factor is $-i\Gamma_i$ for either orientation of the fermion flow~\cite{Denner:1992vza}. For $X_i(p_1)X_j(p_2)\to\mathcal F(k_1)\mathcal F(k_2)$,
\begin{equation}
\begin{aligned}
\mathcal M={}&\frac{[\bar u(k_1)\Gamma_iu(p_1)][\bar u(k_2)\Gamma_ju(p_2)]}{t-M_\sigma^2}\\
&-\frac{[\bar u(k_2)\Gamma_iu(p_1)][\bar u(k_1)\Gamma_ju(p_2)]}{u-M_\sigma^2},
\end{aligned}
\end{equation}
where the relative sign follows from the odd permutation of external spinors. Helicity configurations $(-,-)$, $(+,+)$ and mixed correspond to $ff$, $\bar f\bar f$ and $f\bar f$, respectively. A symmetry factor $1/2$ accounts for the identical final Majorana states, and initial spins are averaged. For $\sigma(p_1)\sigma(p_2)\to\mathcal F(k_1)\mathcal F(k_2)$,
\begin{equation}
\mathcal M=\sum_{i=\pm}\bar u(k_1)\,\Gamma_i\left[\frac{\slashed q_t+M_i}{q_t^2-M_i^2}+\frac{\slashed q_u+M_i}{q_u^2-M_i^2}\right]\Gamma_i\,v(k_2),
\end{equation}
with $q_t=k_1-p_1$ and $q_u=k_1-p_2$. The projections $P_R(\slashed q+M)P_R=MP_R$ and $P_R(\slashed q+M)P_L=P_R\slashed q$ make explicit that $ff$ ($\bar f\bar f$) is proportional to $\sum_ig_i^2M_i$ ($\sum_ig_i^{*2}M_i$), and therefore to $\mathcal D$ in Eq.~\eqref{eq:ss}, while $f\bar f$ involves $|g_i|^2\slashed q$. All spin sums were evaluated numerically with explicit chiral-basis spinors, $v=C\bar u^T$, and Gauss--Legendre angular integration. The results reproduce Eqs.~\eqref{eq:pwave}--\eqref{eq:ss} and pass the Dirac-limit checks of Sec.~\ref{sec:xsec}.

\section{Modal-charge conservation in the localized theory}\label{app:modal}
For each mode define $q(\bm n)=\bm n$ with addition modulo two. The overlap condition $\ell=m\oplus n$ is equivalent to
\begin{equation}
q(S_\ell)+q(F_m)+q(F_n)=0\quad\text{in }(\mathbb Z_2)^3 .
\end{equation}
In any connected diagram built from these cubic vertices and the Majorana mass insertion (which preserves $q$), summing this relation over all vertices counts each internal line twice. Every internal charge therefore cancels modulo two, and $\bigoplus_{\rm external}q=0$. An initial $X_-$ carries $q=100$ and cannot decay into states composed solely of $000$ modes. At the benchmark the lightest state with $q_1=1$ other than $X_\pm$ is $S_{100}$, which lies above $X_-$, and the first continuum threshold is higher still. This establishes stability within the localized interaction basis. It does not exclude ultraviolet operators that violate the selection rule.

\section{Numerical procedure}\label{app:num}
The thermal averages use the relativistic expression for unequal masses~\cite{Gondolo:1990dk,Edsjo:1997bg},
\begin{equation}
\langle\sigma_{ij}v\rangle=\frac{\int ds\;\sigma_{ij}(s)\,\lambda(s,M_i^2,M_j^2)\,K_1(\sqrt s/T')/\sqrt s}{8M_i^2M_j^2\,T'\,K_2(M_i/T')K_2(M_j/T')},
\end{equation}
where $s$ is integrated from $(M_i+M_j)^2$ and $\sigma_{ij}(s)$ is tabulated from the exact tree-level amplitudes. The SM energy and entropy densities are computed for an ideal gas of SM particles with physical masses, with a smooth interpolation to a hadron gas at $T_c\simeq170\MeV$; this gives $g_*=106.75$ above the electroweak scale, $76$ at $1\GeV$ and $10.75$ before $e^\pm$ annihilation. The dark temperature follows from Eq.~\eqref{eq:R} with full Fermi--Dirac and Bose--Einstein integrals. Equation~\eqref{eq:boltz} is integrated in $\ln T$ with a stiff solver from $x'=3$, where $Y=Y_{\rm eq}$, down to $T=2\MeV$. As a validation, the solver reproduces $\lh=5.39$ when restricted to the $f\bar f$ channels of Eqs.~\eqref{eq:pwave}--\eqref{eq:mixedffbar}, two states, $g_*=80$ and a common temperature, i.e.\ the standard treatment of a Majorana fermion with a complex $t$-channel mediator, and agrees with an independent implementation of that treatment to $10^{-4}$.


\end{document}